\documentclass[12pt,aps,prb,preprint]{revtex4} 

\usepackage{amsmath}

\newif\ifpdf
  \ifx\pdfoutput\undefined
    \pdffalse
  \else
    \pdfoutput=1
  \pdftrue
\fi

\ifpdf
  \usepackage[pdftex]{graphicx}
\else
  \usepackage{graphicx}
\fi

\begin{document}

\title{The motion of the freely falling chain tip}
\author{W. Tomaszewski}
\email{waltom@phys.put.poznan.pl} 

\author{P. Pieranski}
\email{pieransk@man.poznan.pl} 
\affiliation{Poznan University of Technology, Poland}

\author{J.-C. Geminard}
\email{jean-christophe.geminard@ens-lyon.fr}
\affiliation{Ecole Normale Sup\'erieure, Lyon, France}

\date{October 5, 2005}

\begin{abstract}

The dynamics of the tip of the falling chain is analyzed. Results of laboratory experiments are presented and compared with results of numerical simulations. Time dependences of the velocity and the acceleration of the chain tip for a number of different initial conformations of the chain are determined. A simple analytical model of the system is also considered. 

\end{abstract}

\maketitle

\section{Introduction}

The problem of bodies falling in a gravitational field is so old that hardly anything new could be expected to add to it. However, the development of the numerical simulation methods has opened for the analysis a few interesting cases, difficult to analyze with analytical means. The dynamics of a falling chain is among them. 

There are a few variations of the problem. For instance, one can consider here the chain initially gathered in a compact heap located on a table, close to one edge. The motion starts when one of the chain ends is brought over the edge. If one assumes that the chain leaves the heap without friction, the problem becomes tractable analytically and, as one can demonstrate, it falls with a constant acceleration. The surprise is that the acceleration is not $g$, as one would expect, but $g/3$ \cite{SousaRodrigues2004}.

In the variation of the falling chain problem considered here the chain is initially attached at both ends to a horizontal support. Then, as one of the ends is released, the chain starts falling. The case when the initial horizontal distance $\Delta x$ between the ends is small in the comparison to the chain length $L$ has been considered before \cite{Schagerl1997}. We repeated the experiments demonstrating that, in this case, the end of the chain falls with an increasing acceleration, always greater than $g$; this apparently paradoxical result can be explained via analytical calculations \cite{TomPier2005}. In the present paper, we provide the detailed analysis of the problem and we extend the study to the case of large $\Delta x$.

What happens when the initial horizontal separation $\Delta x$ of the chain ends increases, in particular when it reaches its largest value $\Delta x=L$, was not known. We describe in what follows a series of laboratory experiments performed at different horizontal separations between the ends of the chain and compare the measurements with the results of numerical simulations. The not studied before case, in which the initial distance between the ends of the chain equals $L$ (i.e. when the chain is initially maximally stretched) proves to be very interesting.

\section{The fall of the tightly folded chain -- Analytical solution}

\begin{figure}[htbp]
	\begin{center}
		\includegraphics[scale=0.3]{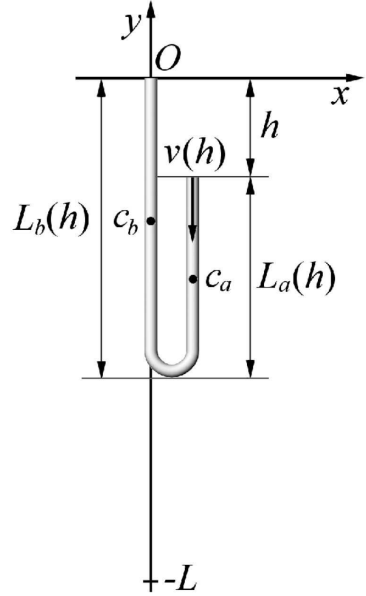}
	\end{center}
	\caption{\label{Figure001}Geometry of the conformation of the tightly folded chain at time $t>0$. The position of the freely falling chain is described in terms of $h$. Part $a$) of the chain is falling down while part $b$) is motionless; we denote by $c_a$ and $c_b$ their centers of mass.}
\end{figure}

To get an intuitive insight into what we can expect in the experiments with the falling chain, let us consider first the case $\Delta x=0$ (figure \ref{Figure001}). We introduce here an analytically tractable simplified model, in which we assume that the conformations explored by the falling chain can be always seen as consisting of two pieces: $a$) the falling piece that shortens with time; $b$) the almost motionless piece that elongates with time. Such a division of the chain is possible when the initial horizontal separation of the chain ends equals zero and the chain consists of infinitely many, infinitely thin segments. In this limit the chain can be seen as a folded perfectly floppy and infinitely thin continuous filament.

Initially both ends of the chain are attached to a point of support $O$, the vertical position of which $y = 0$. Then, at time $t=0$, one of the ends of the chain is released and the chain starts falling. Figure \ref{Figure001} presents schematically the geometry of the system. 

The dynamics of this simple model can be solved analytically by applying the law of energy conservation \cite{TomPier2005}. We assume that the chain has total length $L$ and that its mass $M$ is distributed uniformly along it. To simplify final analysis of the results, we introduce a new variable $h$ describing the distance of the freely falling tip to its initial position. The $h$-axis is thus oriented downwards, in the direction of the gravitational field. In what follows, we shall refer to $h$ as the \textit{fall distance}. In terms of $h$, the length and mass of the falling part are given by:
\begin{eqnarray}
  L_a(h) = \frac{L-h}{2}, \quad 
  M_a(h) = \frac{(L-h)M}{2L}.
\end{eqnarray} 
The length and mass of the motionless part are: 
\begin{eqnarray}
  L_b(h) = \frac{L+h}{2}, \quad 
  M_b(h) = \frac{(L+h)M}{2L}
\end{eqnarray}
and the corresponding vertical positions of their centers of mass are given by:
\begin{eqnarray}
  y_{c_a}(h) = -h - \frac{L_a(h)}{2}=-\frac{3h+L}{4}, \quad 
  y_{c_b}(h) = -\frac{L_b(h)}{2}=-\frac{h+L}{4}.
\end{eqnarray} 
The potential energy of the falling part of the chain expressed in terms of the fall distance $h$, relative to the point $y=-L/2$, equals:
\begin{eqnarray}
  U_a(h)=M_a(h)g\left(y_{c_a}(h)+\frac{L}{2}\right)=\frac{Mg(L^2-4Lh+3h^2)}{8L}. 
\label{Ua}
\end{eqnarray} 
Analogously, the potential energy of the motionless part of the chain is given by:
\begin{eqnarray}
  U_b(h)=M_b(h)g\left(y_{c_b}(h)+\frac{L}{2}\right)=\frac{Mg(L^2-h^2)}{8L}. 
\label{Ub}
\end{eqnarray} 
The total potential energy of the chain is thus given by the formula:
\begin{eqnarray}
 U(h)=\frac{Mg(L-h)^2}{4L}. 
\end{eqnarray}
Note that, according to the chosen reference, $U(L)=0$ (i.e. the potential energy is zero when the tip of the chain reaches its lowest position).

The kinetic energy of the falling part of the chain is given by:
\begin{eqnarray}
  T_a(h)=\frac{M_a(h) v_c(h)^2}{2}=\frac{M(L-h)v_c(h)^2}{4L}.
\label{Ta}
\end{eqnarray}
Assuming that the free end of the chain is initially located at $h=0$, we may formulate the law of energy conservation for the falling chain (remember that $T_b=0)$:
\begin{eqnarray}
  U_a(0)+U_b(0) = U_a(h) + T_a(h)+ U_b(h),
\end{eqnarray}
which in view of \eqref{Ua}, \eqref{Ub}, and \eqref{Ta} gives:
\begin{eqnarray}
  \frac{1}{4}MgL = \frac{M(L-h)(gL-gh+v_c(h)^2)}{4L}.
\end{eqnarray}

After straightforward simplifications, we find the formulae describing the velocity $v_c$ and time $t_c$ of the chain tip versus the fall distance $h$:
\begin{equation}
\label{eq14}
  v_c(h)=\sqrt{\frac{{gh\left(2L-h\right )}}{L-h}}, \quad 
  t_c(h)=\int\limits_0^h{\frac{ds}{v_c(s)}}=\int\limits_0^h{\sqrt{\frac{L-s}{{gs\left(2L-s\right )}}}}ds.
\end{equation}

\begin{figure}[htbp]
	\begin{center}
		\includegraphics[scale=0.35,angle=-90]{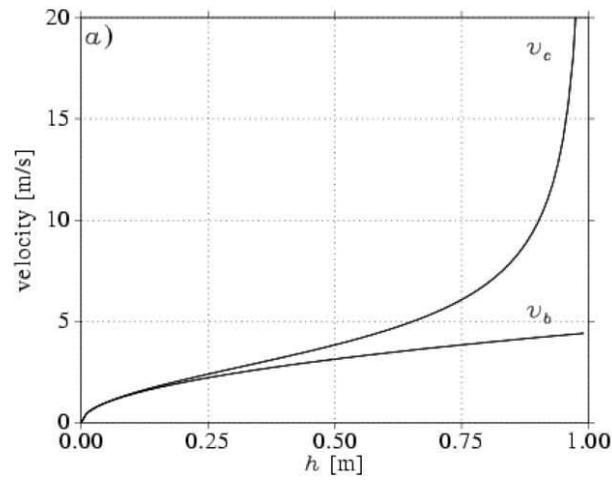}
		\includegraphics[scale=0.35,angle=-90]{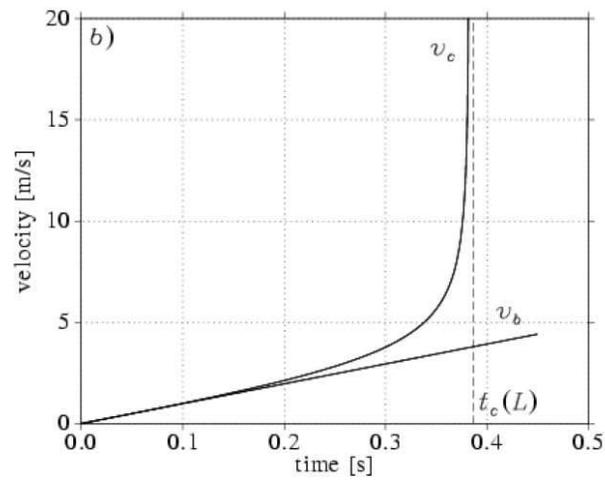}
	\end{center}
	\caption{\label{Figure002} The velocity of the falling chain tip $v_{c}$ and the compact body $v_{b}$ versus: $a$) the fall distance $h$ and $b$) time $t$.}
\end{figure}

\begin{figure}[htbp]
	\begin{center}
		\includegraphics[scale=0.35,angle=-90]{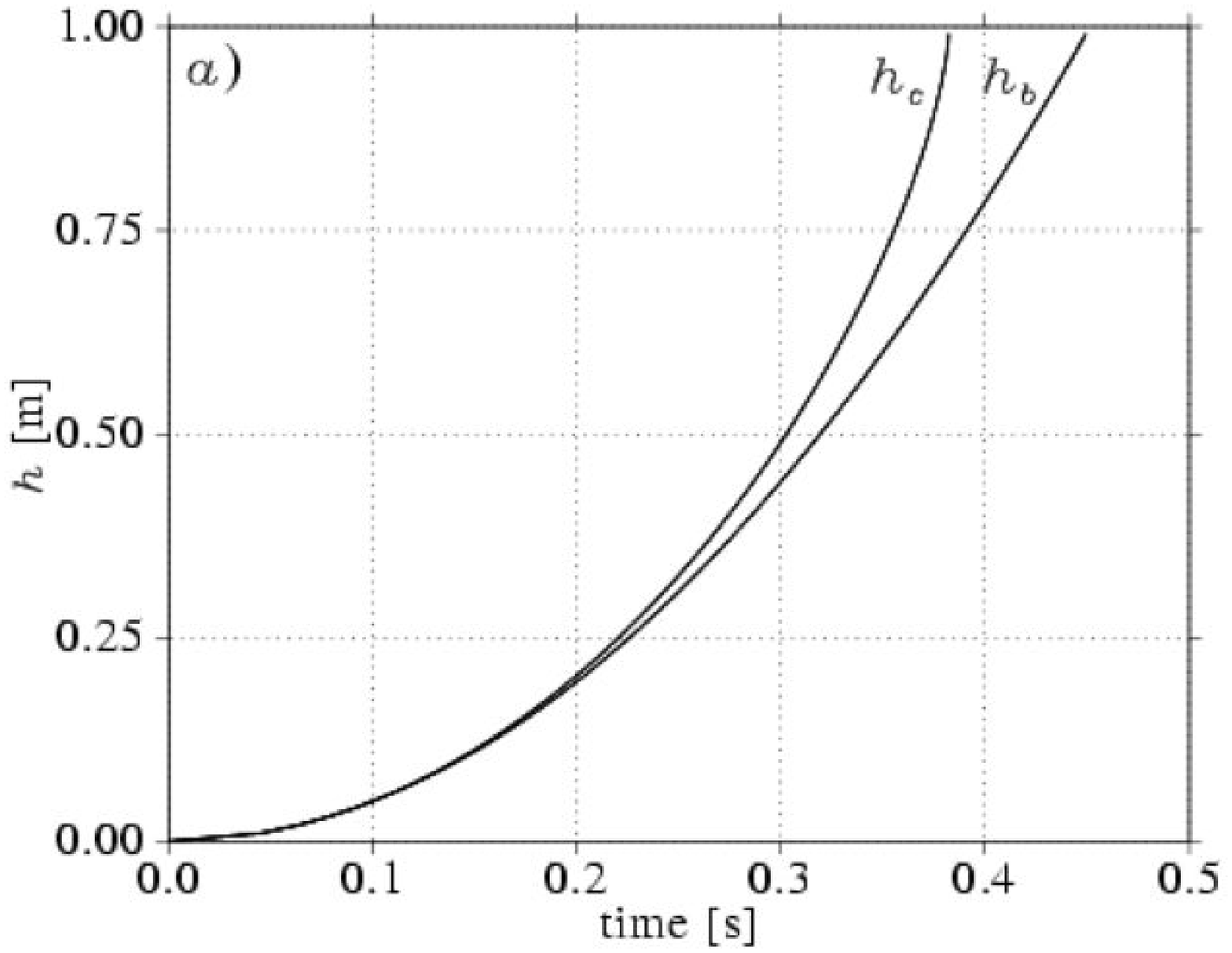}
		\includegraphics[scale=0.35,angle=-90]{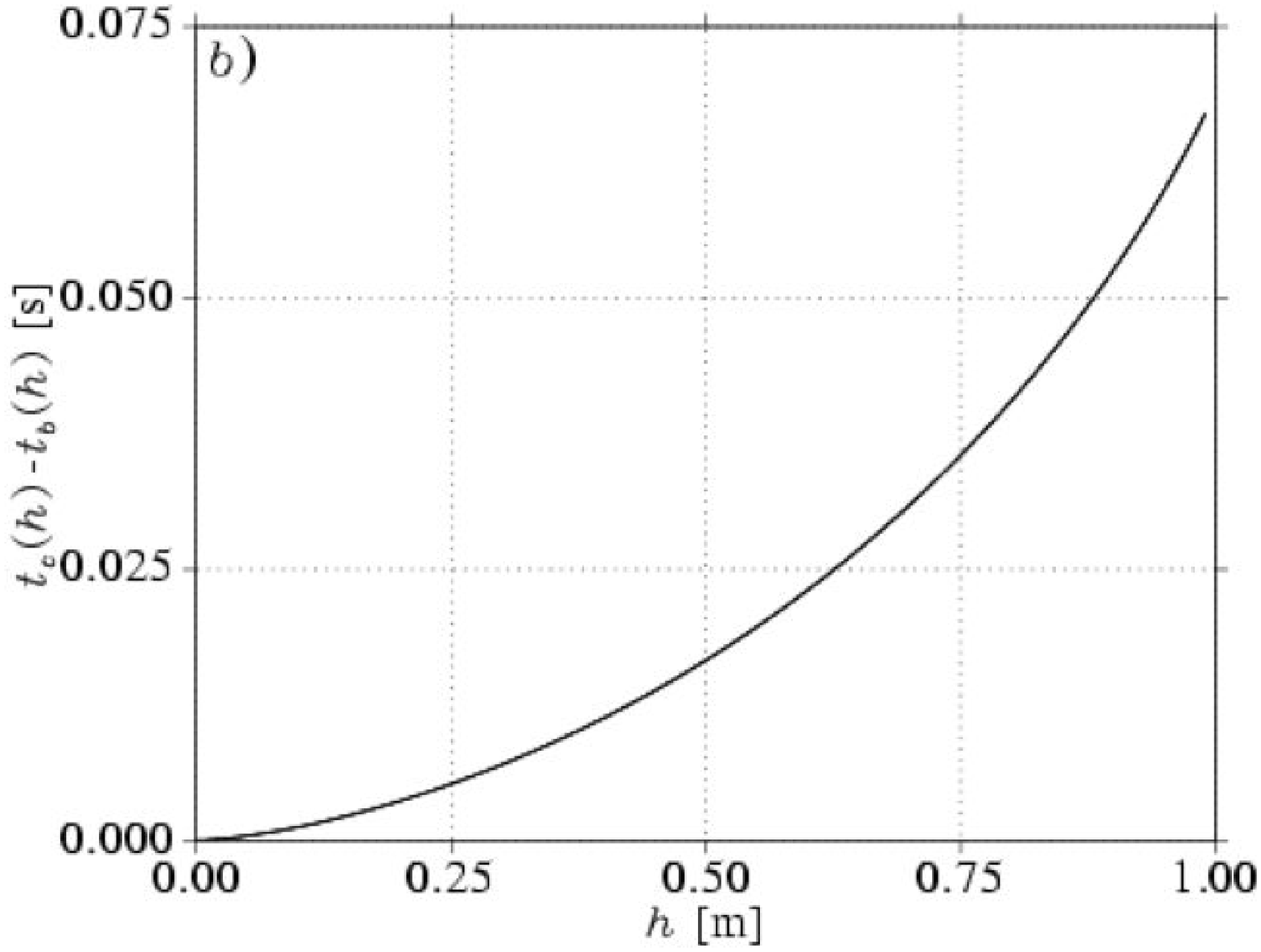}
	\end{center}
	\caption{\label{Figure003}$a$) The fall distance $h_c$ of the tip of the chain and the compact body, $h_b$ versus time $t$. $b$) The difference between times at which the compact body and the chain tip reach the same fall distance $h$.}
\end{figure}

Let us compare the motion of the freely falling chain and that of a compact body. Formulae describing the motion of the compact body can be also derived from the energy conservation law:
\begin{equation}
 \label{15}
 g\left(\frac{1}{2}L\right)=g\left(\frac{1}{2}L+h\right)+\frac{1}{2}v_b(h)^2.
\end{equation}
Solving equation \eqref{15}, we find the velocity $v_b$ and time $t_b$ of the freely falling compact body in terms of the fall distance $h$:
\begin{eqnarray}
  v_b(h)=\sqrt{2gy}, \quad t_b(h)=\int\limits_0^h{\frac{ds}{\sqrt{2gs}}}=\sqrt{2h/g}.
\end{eqnarray}
Figures \ref{Figure002} and \ref{Figure003} present a comparison between the dynamics of the folded chain and the compact body calculated for $L = 1$~m and the gravitational acceleration $g = 9.81$~m/s$^2$.

Figure \ref{Figure002}$a$ shows that the velocity of the chain tip is, at any fall distance $h$, larger than the velocity of the compact body. As expected (figure \ref{Figure002}$b$), the acceleration of the falling body is constant and equals $g$ while that of the falling chain tip increases with time, always exceeding $g$.

Comparing the motions of the falling chain and of the compact body one may ask about the difference between the times at which they reach the same fall distance $h$. Figure \ref{Figure003} presents plots providing a clear answer to this question.

The analytical results obtained above for the simplified model of the falling chain may be confronted with results of laboratory experiments performed on a real chain. The results obtained for various initial configurations are the subject of the next section \ref{sec:LaboratoryExperiments}.

\section{Laboratory experiments}
\label{sec:LaboratoryExperiments}

 We are aiming at the experimental study of the falling-chain dynamics and, more specifically, at its comparison with the dynamics of a freely falling weight. In order to point out the differences in their trajectories, we designed an experimental setup that makes it possible to record the simultaneous motions of the two objects.

\begin{figure}[htbp]
	\begin{center}
		\includegraphics[scale=0.4]{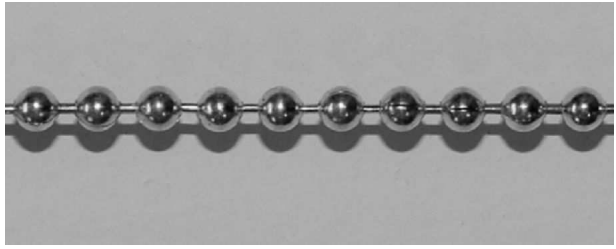}
	\end{center}
	\caption{\label{Figure004}Stainless chain used in the laboratory experiments.}
\end{figure}

The chain consists of stainless-steel identical segments which are made from rods and spheres attached to each other (figure \ref{Figure004}). The total length of a segment $l = (4.46 \pm 0.01)10^{-3}$~m and the diameter of the spheres $\phi = (3.26 \pm 0.01)10^{-3}$~m. In addition, we determine the minimum radius of curvature $R_{min} = (4.8 \pm 0.2)~10^{-3}$~m, which the chain can present without loading any elastic energy. We use a chain of length $L = 1.022$~m, which corresponds to $n = 229$ segments for a total mass $M = (2.08 \pm 0.01)10^{-2}$~kg.

\begin{figure}[htbp]
	\begin{center}
		\includegraphics[scale=0.4]{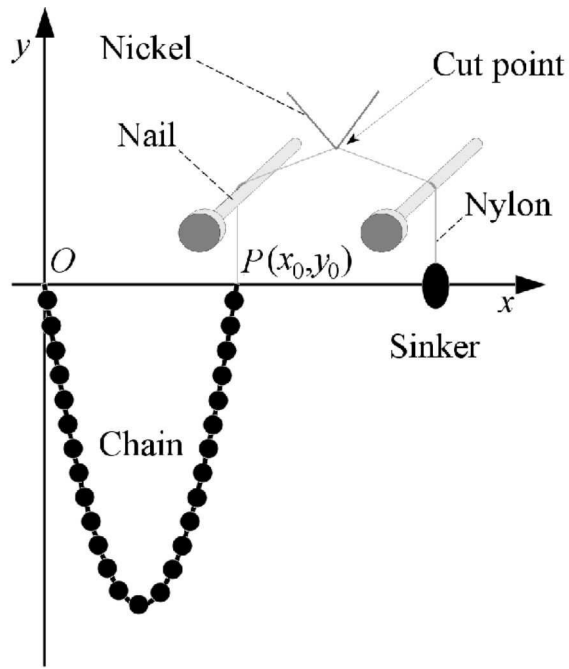}
	\end{center}
	\caption{\label{Figure005}Sketch of the experimental setup.}
\end{figure}

The chain is tightly attached at one end to a firm support $O$ by means of a thin thread. See figure \ref{Figure005}. At the other edge located at point $P=(x_0, y_0)$, the chain ends with a rod (we open and remove the last sphere) to which we tie a thin nylon wire (fishing line, diameter $0.1~10^{-3}$~m). The free-falling weight, a sinker (a lead weight used for sinking fishing lines) of mass $\mathcal{M} = 10^{-2}$~kg, is then attached to the other end of the nylon wire (length about 5~cm). We then make the nylon wire hang to two nails and a thin metallic wire (nickel, diameter $10^{-4}$~m) as sketched in figure \ref{Figure005}.
The whole system is adjusted so as to insure that the sinker and the end of the chain are at the same altitude as the other end of the chain in $O$ ($y=0$). It can be displaced horizontally in order to choose the initial horizontal separation between the two ends of the chain. As the mass of the sinker $\mathcal{M}$ is about half the total mass of the chain $M$, the system is almost always equilibrated (in addition, the equilibrium is helped by the solid friction in the contact regions of the nylon wire with the nails and the metallic wire). Thus, the initial conformation that is formed by the chain after damping of all the oscillations, is close to the catenary curve
\footnote{
A word of caution seems here necessary; from the rigorous point of view the initial conformation formed in the experiments with chains built from a finite number of segments is not a catenary curve. That this is the case one can see immediately considering the conformation of a chain built from an odd number of segments whose ends are fixed to points located at the same level and separated in the horizontal direction by the length of a single segment. In such a case the initial conformation consists of two exactly vertical pieces. This was the conformation used by Schagerl in his laboratory experiments \cite{Schagerl1997}}.

Injecting a large electric current (about 1~A) in the nickel wire results in cutting suddenly the nylon wire at the point where they cross; the weight and the end of the chain then simultaneously start falling freely under the action of gravity. We point out that they both fall with a small piece of nylon wire attached to them. However, as the force that pushes the wire against the nails vanishes, the friction force vanishes as soon as the wire is cut. In addition, during the free fall, the pieces of wire have no effect on the dynamics as the mass of nylon is negligible in comparison to the mass of the sinker or chain.

\begin{figure}[htbp]
	\begin{center}
		\includegraphics[scale=0.6]{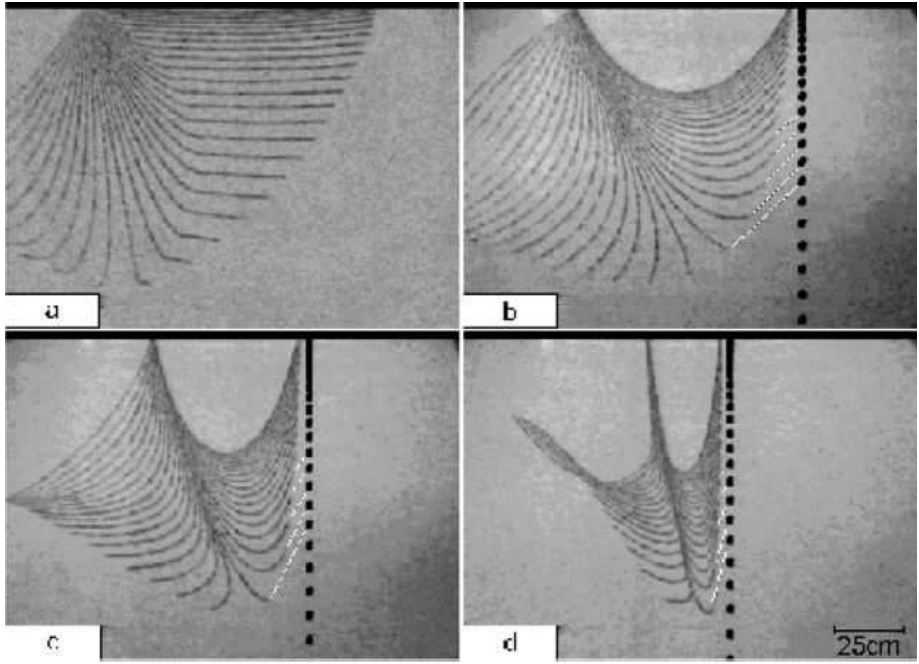}
	\end{center}
	\caption{\label{Figure006}Successive conformations of the falling chain vs. time. The left end of the chain remains attached to the frame, while the right end is free to fall due to gravity. In $b$), $c$) and $d$), the white straight-lines connect the free falling end to the free falling weight for the last five images before the maximum extension of the chain (length $L = 1.022$~m, time spacing between the successive images 1/50 s, initial separation between the chain ends: $a$) $x_0=1.019$~m, $b$) $x_0=0.765$~m, $c$) $x_0=0.510$~m, and $d$) $x_0=0.255$~m).}
\end{figure}

The falling chain and weight are imaged with the help of a standard CCD video camera (Panasonic WV-BP550/G) and the images are recorded on a video cassette recorder. The chosen shutter speed ($1/4000$~s) is enough for obtaining clear images of both the chain and sinker (figure \ref{Figure006}). The movies are digitized afterwards by means of a Macintosh computer equipped with a frame grabber board (Data Translation DT2255). Further analysis with an image-processing software (NIH-Image) makes possible to recover 50 images per second from the movies which are initially made from 25 interlaced images per second. The interlacing allows thus to double the time resolution but results in a loss in the spatial resolution, which is typically of about 4~mm per pixel.

The positions of both the falling chain tip and the weight at the times $t_i, i=0,1,2,...$, at which consecutive frames were recorded, is determined from the digitized images. To make the discussion of the results as simple as possible, the experimentally determined positions of the falling object will be given as the vertical $h$ and horizontal $w$ deviations of its current $x(t)$ and $y(t)$ coordinates from their initial values $(x_0,y_0)$: 
\begin{equation}
\begin{split}
	w(t) &= x_0-x(t), \\
	h(t) &= y_0-y(t), 
\end{split}
\end{equation}
In what follows we shall refer to the variables as the vertical, $h$, and horizontal, $w$, fall distances. According to their definitions, in the initial stages of the falling process both of the falling distances are positive. In all experiments $y_0=0$, while $x_0$ was changing in four steps from $1$~m to $0.25$~m. Note that since the motionless end of the chain is attached to point $(0,0)$, the initial horizontal separation of the chain ends $\Delta x = x_0$. In view of this equality, in the following we shall be denoting the initial separation by $x_0$. Results of the laboratory experiments will be confronted in section \ref{sec:QuantitativeAnalysis} with results of the numerical simulations presented in section \ref{sec:NumericalExperiments}.

\section{The model of the chain and its equations of motion}
\label{sec:TheModelOfTheChain}

One can define a few discrete models of the chain; below we present one of them. Its equations of motion will be formulated for the case, in which one of the chain ends is attached to the fixed support while the other one is free. Let us note that similar models have been considered before \cite{Schagerl1997, Kommers1995, GalanFraserAchesonChampneys2001}.

\begin{figure}[htbp]
	\begin{center}
		\includegraphics[scale=0.5]{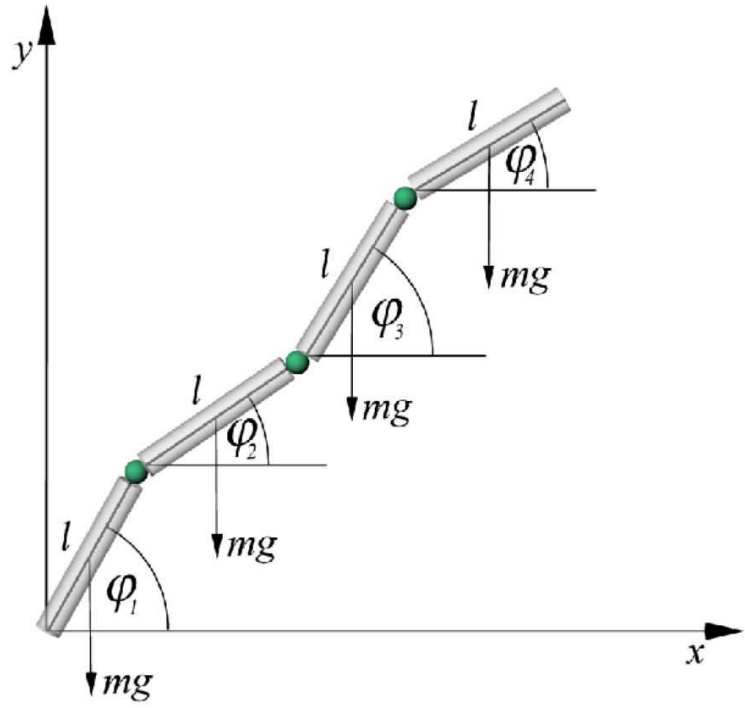}
	\end{center}
	\caption{\label{Figure007}Model of the chain.}
\end{figure}

Consider a chain moving in a gravitational field. Several assumptions will be made to simplify the model. First of all, we assume that the chain is constrained to move only in the vertical plane denoted by $(x, y)$. A chain of mass $M$ and length $L$ consists of $n$ thin cylindrical rods (in the following we shall refer to them as segments) with masses $m_i = m = M/n$, $i = 1..n$, and lengths $l_i = l = L/n$, $i = 1..n$. All the segments are rigid and cannot be deformed. Consecutive segments are connected by joints with friction. Figure \ref{Figure007} shows the geometric representation of our model.

In order to formulate the equations of motion, generalized coordinates, which rigorously determine the state of the system, must be specified. Following our predecessors \cite{Kommers1995}, we decide to describe the system using angular coordinates indicating the inclination of the consecutive segments with respect to the $x$-axis.

The position of the first element is determined by the angle $\varphi_{1}$. Similarly, the position of the second element is described by the angle $\varphi_{2}$. The global conformation of the chain in the plane is uniquely expressed by all angles $\varphi_{i}$, $i=1..n$. The angles are below referred to as generalized coordinates of the system. A generalized coordinate $\varphi_{i}$ indicates an angle between the $i$-th element of a chain and the horizontal axis $x$.

The Cartesian coordinates of the $i$-th mass center $(x_i, y_i)$ can be written as follows:
\begin{equation}
\label{eq1}
\begin{split}
  x_{i} &= {\sum\limits_{j=1}^{i-1}{l\cos\varphi_{j}}}+
          \frac{1}{2}l\cos\varphi_{i}, \\
  y_{i} &= {\sum\limits_{j=1}^{i-1}{l\sin \varphi_{j}}} +
          \frac{1}{2}l\sin \varphi_{i}.
\end{split}
\end{equation}

Using the generalized coordinates we shall derive the Lagrange equations of motion. To start with, we shall consider the energy of the system. The motion of the chain is considered as a combination of translational and rotational motions of its segments. Each segment has the moment of inertia $I_{i}=1/12ml^{2}$, $i=1..n$, calculated around the axis perpendicular to the $(x,y)$ plane and passing through the center of mass of the segment. Taking into consideration the relations given in equation \eqref{eq1}, the kinetic energy of the chain is given by:
\begin{eqnarray}
\label{eq2}
  T & = & \frac{1}{2}\sum_{i=1}^{n}\left(m\left(\dot{x}_{i}^{2}+\dot{y}_{i}^{2}\right)+
        I_{i}\dot\varphi_{i}^{2}\right),
\end{eqnarray}
where the dot represents the derivative with respect to the time $t$. The potential energy of the $i$-th segment is given by $mgy_{i}$, where $g$ is the gravitational acceleration. Thus, the potential energy of the chain may be expressed as:
\begin{eqnarray}
\label{eq3}
  U & = & \sum_{i=1}^{n}{mgy_i}.
\end{eqnarray}

To make our model more general, we introduce damping as a Rayleigh's dissipation function \cite{Goldstein1980}:
\begin{eqnarray}
\label{eq4}
  \mathcal{R} = \frac{1}{2}r\sum_{i=1}^{n}\left(\dot\varphi_{i}-\dot\varphi_{i-1}\right)^{2},
\end{eqnarray}
where $r$ is the dissipation coefficient. We assume that the joint which connects the first element of the chain to the support is free of dissipation. This is equivalent to the assumption that $\dot\varphi_{0} = \dot\varphi_{1}$. Similar definition of dissipation was used by other authors \cite{Schagerl1997, GalanFraserAchesonChampneys2001}.

The motion of the falling chain is governed by the system of Lagrange equations of second kind:
\begin{eqnarray}
\label{eq5}
  \frac{d}{dt}\left(\frac{\partial\mathcal{L}}{\partial\dot\varphi_{i}}\right) - 
\frac{\partial\mathcal{L}}{\partial\varphi_{i}}+\frac{\partial\mathcal{R}}{\partial\dot\varphi_{i}}=0, \qquad i=1..n,
\end{eqnarray}
where $\mathcal{L}=T-U$ is the Lagrangian of the system. Applying \eqref{eq2}, \eqref{eq3}, \eqref{eq4} and \eqref{eq5} we find the set of $n$ equations describing the motion of a chain: 
\begin{eqnarray}
\label{eq6}
\sum\limits_{j = 1}^n {m_{i,j}c_{i,j} \ddot\varphi _j } = -\sum\limits_{j = 1}^n {m_{i,j}s_{i,j}\dot\varphi _j^2 } + \frac{r}{ml^2}\left(\dot\varphi_{i-1}-2\dot\varphi_{i}+\dot\varphi_{i+1}\right) 
- \frac{g}{l}a_ic_{i}, \quad i=1..n,
\end{eqnarray}
where $c_{i} = \cos(\varphi_{i})$, $c_{i,j} = \cos(\varphi_{i}-\varphi_{j})$, $s_{i,j} = \sin(\varphi_{i}-\varphi_{j})$, 
$a_{i} = n - i + \frac{1}{2}$ and $m_{i,j}=~\left\{ {\begin{array}{*{20}l}
   {n - i + \frac{1}{3}}, & {i = j}  \\
   {n - \max (i,j) + \frac{1}{2}}, & {i \neq j}
\end{array}}\right.$.

The next section \ref{sec:NumericalExperiments} is dedicated to the results of numerical solving of equation \eqref{eq6}.

\section{Numerical experiments}
\label{sec:NumericalExperiments}

Equations of motion derived in the previous section can be integrated numerically thus allowing one to simulate numerically the motion of the falling chain. In presence of dissipation, the resulting system of equations becomes stiff and requires specific numerical methods. We selected the \texttt{RADAU5} algorithm by Hairer \& Wanner (\url{http://www.unige.ch/~hairer/software.html}) designed for stiff problems. It is based on the implicit Runge-Kutta scheme of order five with the error estimator of order four \cite{HairerWanner1996}.

We performed a series of numerical simulations aiming to reproduce results of the experiments described in section \ref{sec:LaboratoryExperiments}. Thus, as the initial configuration of the chain we used the discrete catenary curve shown in figure \ref{Figure005} with four different separations between the ends of the chain: $a$) $x_0=1.019$~m, $b$) $x_0=0.765$~m, $c$) $x_0=0.510$~m and $d$) $x_0=0.255$~m identical with the separations used in the laboratory experiments. Numerical simulations were performed with $n=229$, $L=1.02$~m, $M=0.0208$~kg, $g=9.81$~m/s$^2$ and time $t \in [0,0.5]$~s. The only free parameter left was thus the dissipation parameter $r$. Varying it we aimed at obtaining the best agreement of the numerical results with the laboratory experiments. To compare the results, we monitored the distance between the positions of the chain tip found in the consecutive frames of the video recordings and the positions found in the numerical simulations at the same times. The distance between laboratory and numerical data obtained in a single experiment is defined as follows:
\begin{eqnarray}
	\delta = \sqrt{\frac{1}{N}\sum_{i=1}^{N} (w_i- \hat w_i)^2+(h_i- \hat h_i)^2},
\end{eqnarray}
where the $N$ denotes the number of analyzed frames. Points $(w_i, h_i)$ and $(\hat w_i, \hat h_i)$ for $i=1..N$ are here the horizontal and vertical deviation from the initial position of the chain tip found in consecutive frames of the laboratory and numerical experiments, respectively. In order to find the optimal value of $r$ providing the best fit for all four experiments $a)$, $b)$, $c)$ and $d)$ we determined the total distance
\begin{eqnarray}
	\Delta = \delta^{(a)}+\delta^{(b)}+\delta^{(c)}+\delta^{(d)}.
\end{eqnarray}
$\Delta$ depends on the assumed value of $r$; we have found its values for $r$ in the range from $10^{-6}$ to $10^{-4}$. Then, $\Delta(r)$ was analyzed with the use of the \textit{least-square} algorithm based on the procedure \texttt{SVDFIT} \cite{PressFlanneryTeukolskyVetterling1992}. The optimal value of dissipation parameter was found to be equal $r=2.163\cdot10^{-5}$~Nms (i.e. for this one, the $\Delta(r)$ reaches its minimum value equal to $0.02510$~m).

It seems interesting to check how this single value of $r$ fits the data obtained in each of the four experiments; table \ref{tab1} shows the results.

In all cases, $\delta$ are relatively small until the chain reach its minimal vertical position (it is less then $0.004$~m). It becomes much bigger after the chin tip start to raise and it has the great influence in the value of $\Delta$. Figure \ref{Figure008} presents the experimental vertical $h$ and horizontal $w$ fall distances together with their numerical counterparts determined with a much smaller time step.

Consecutive conformations of the falling chain found in the numerical simulations are presented in figure \ref{Figure009}. The conformations correspond to the same times at which they were recorded in the laboratory experiments. Positions of the falling compact body are also shown in the figure. As one can see comparing figures \ref{Figure006} and \ref{Figure009}, the shapes of the experimental and numerical conformations are almost identical.

\section{Quantitative analysis}
\label{sec:QuantitativeAnalysis}

\begin{figure}[htbp]
	\begin{center}
		\includegraphics[scale=0.30,angle=-90]{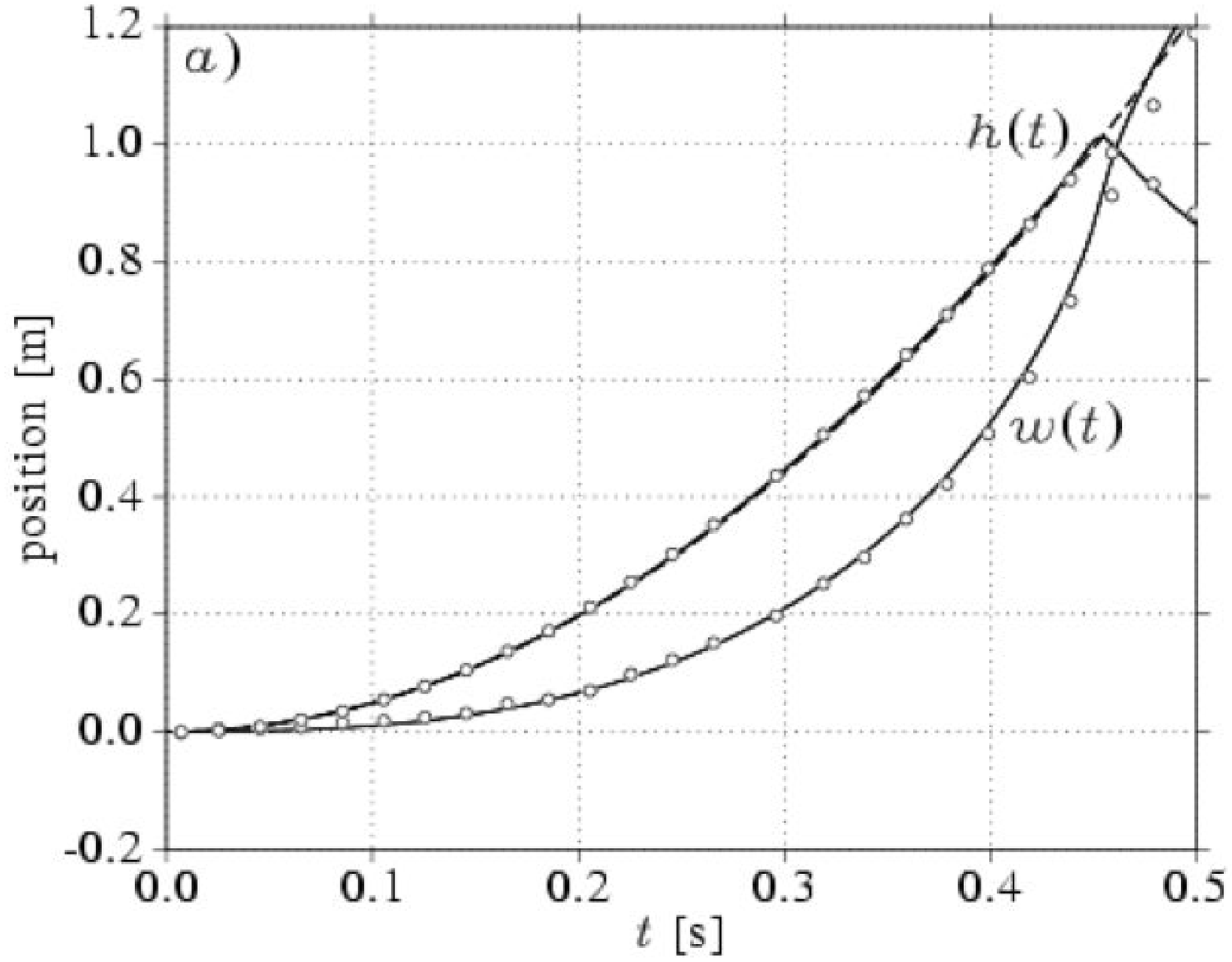}
		\includegraphics[scale=0.30,angle=-90]{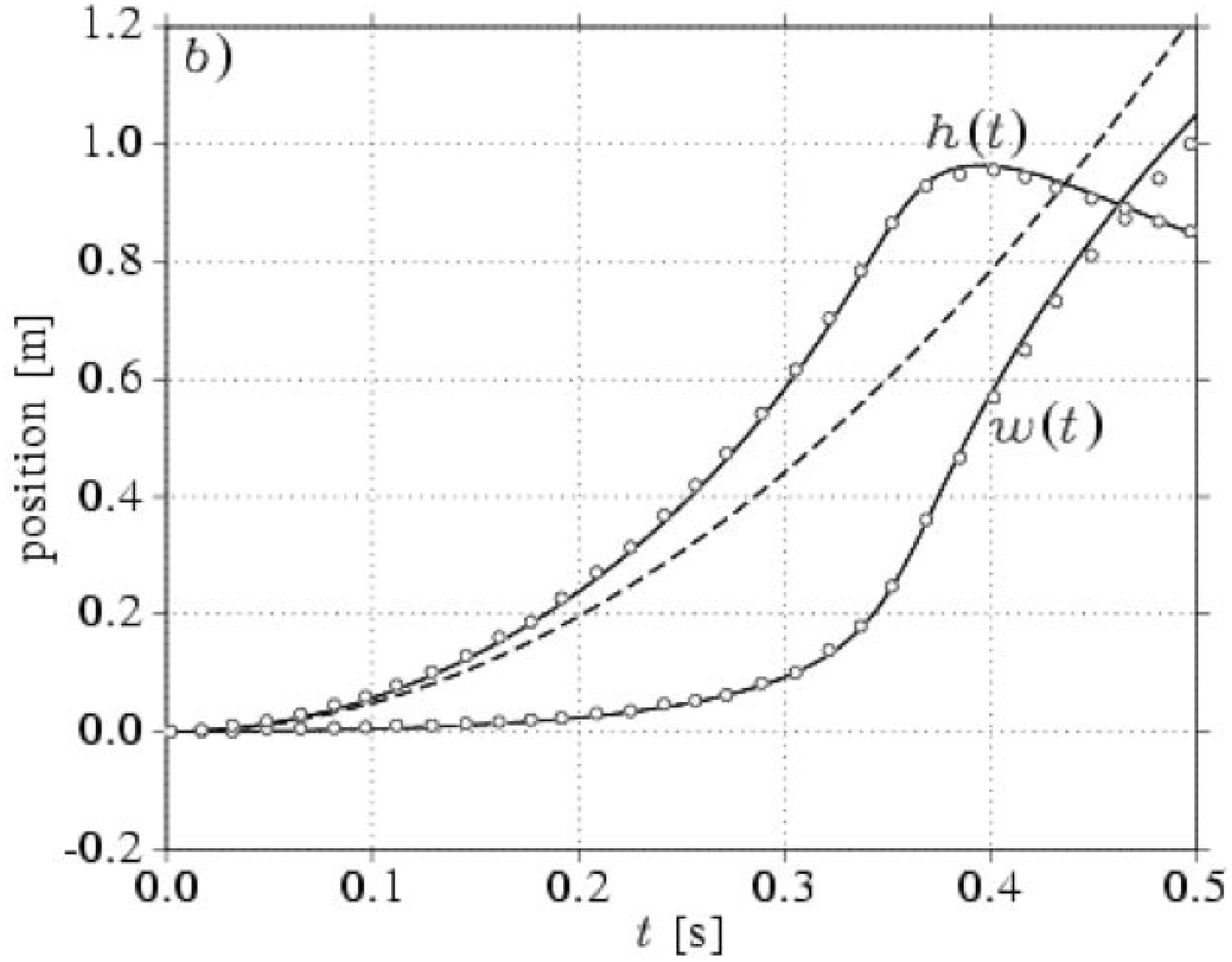}
		\includegraphics[scale=0.30,angle=-90]{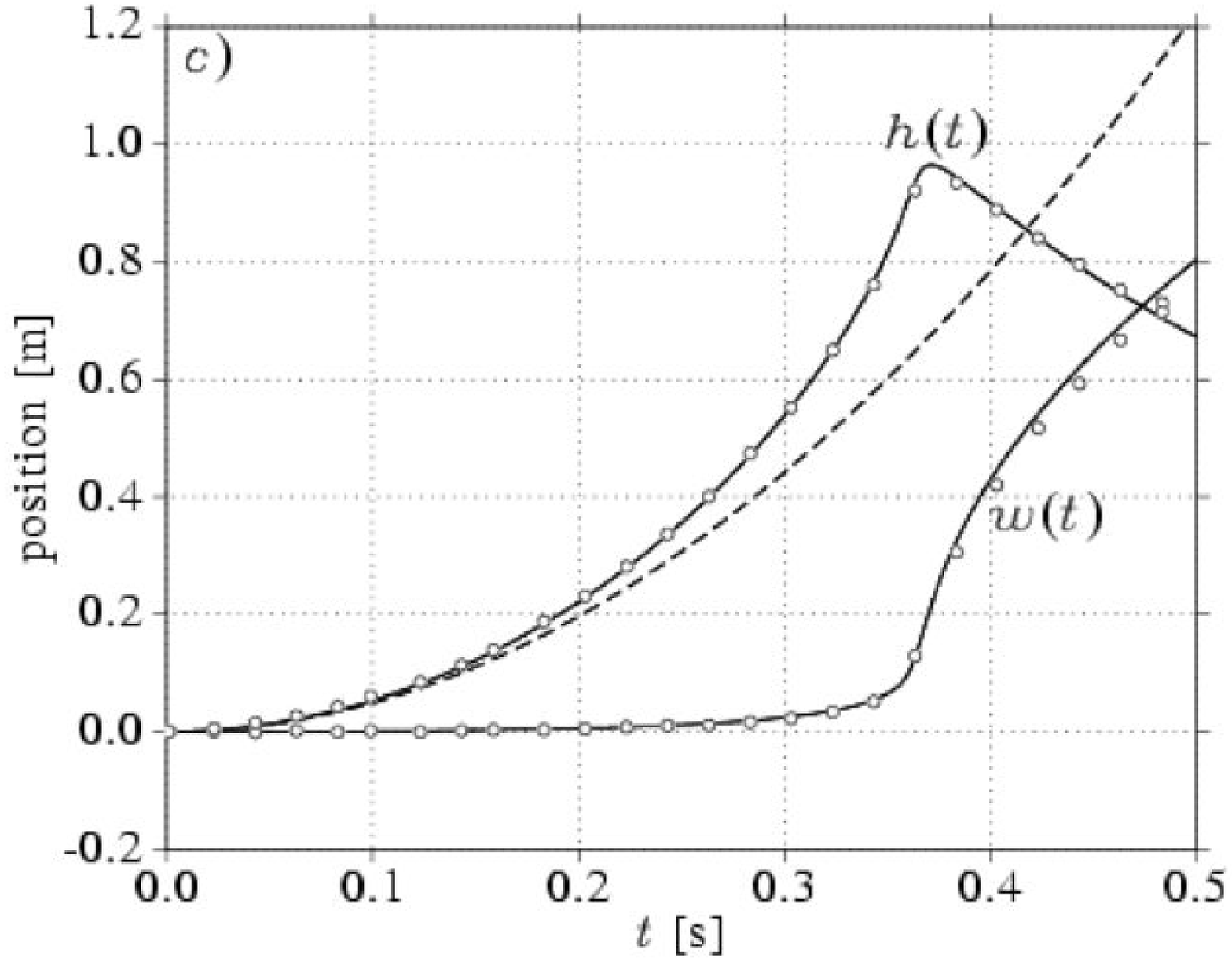}
		\includegraphics[scale=0.30,angle=-90]{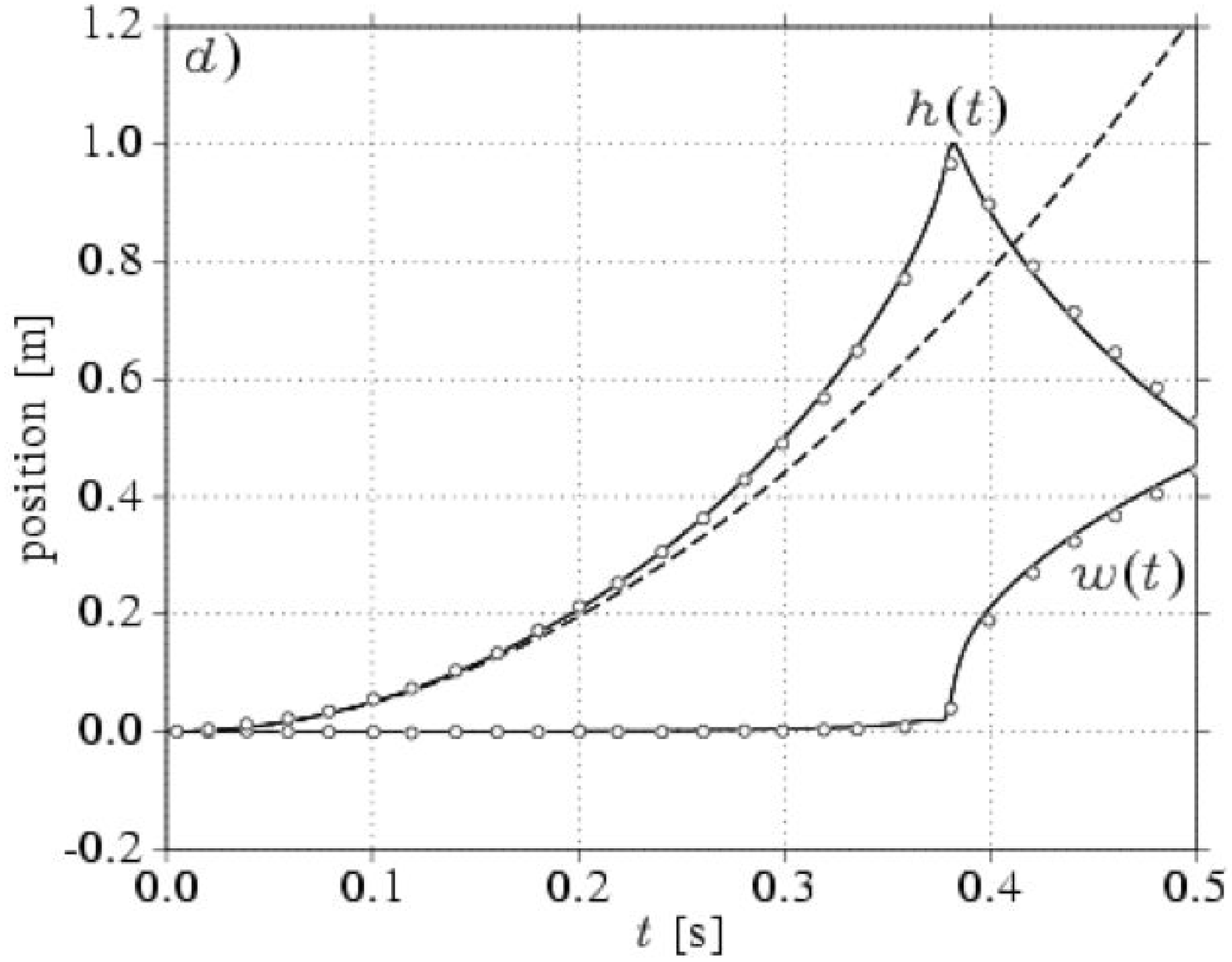}
	\end{center}
	\caption{\label{Figure008}The comparison of the vertical $h$ and horizontal $w$ fall distances of the falling chain tip found experimentally (circles) and numerically (solid lines). The parabola of the compact body fall is also shown (dotted lines). The initial separation between the chain ends: $a$) $x_0=1.0195$~m, $b$) $x_0=0.765$~m, $c$) $x_0=0.51$~m, and $d$) $x_0=0.255$~m.}
\end{figure}

\begin{figure}[htbp]
	\begin{center}
		\includegraphics[scale=0.30,angle=-90]{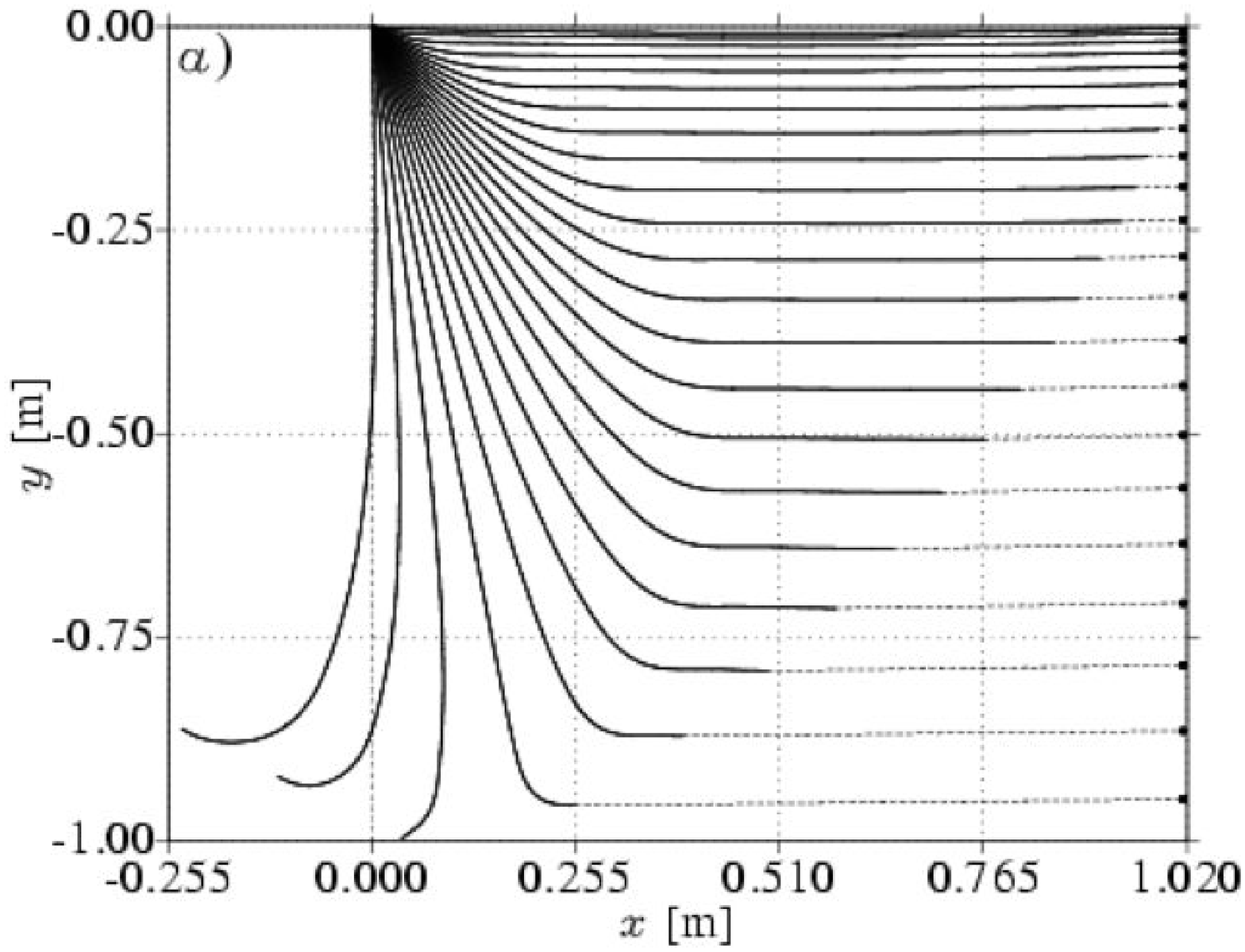}
		\includegraphics[scale=0.30,angle=-90]{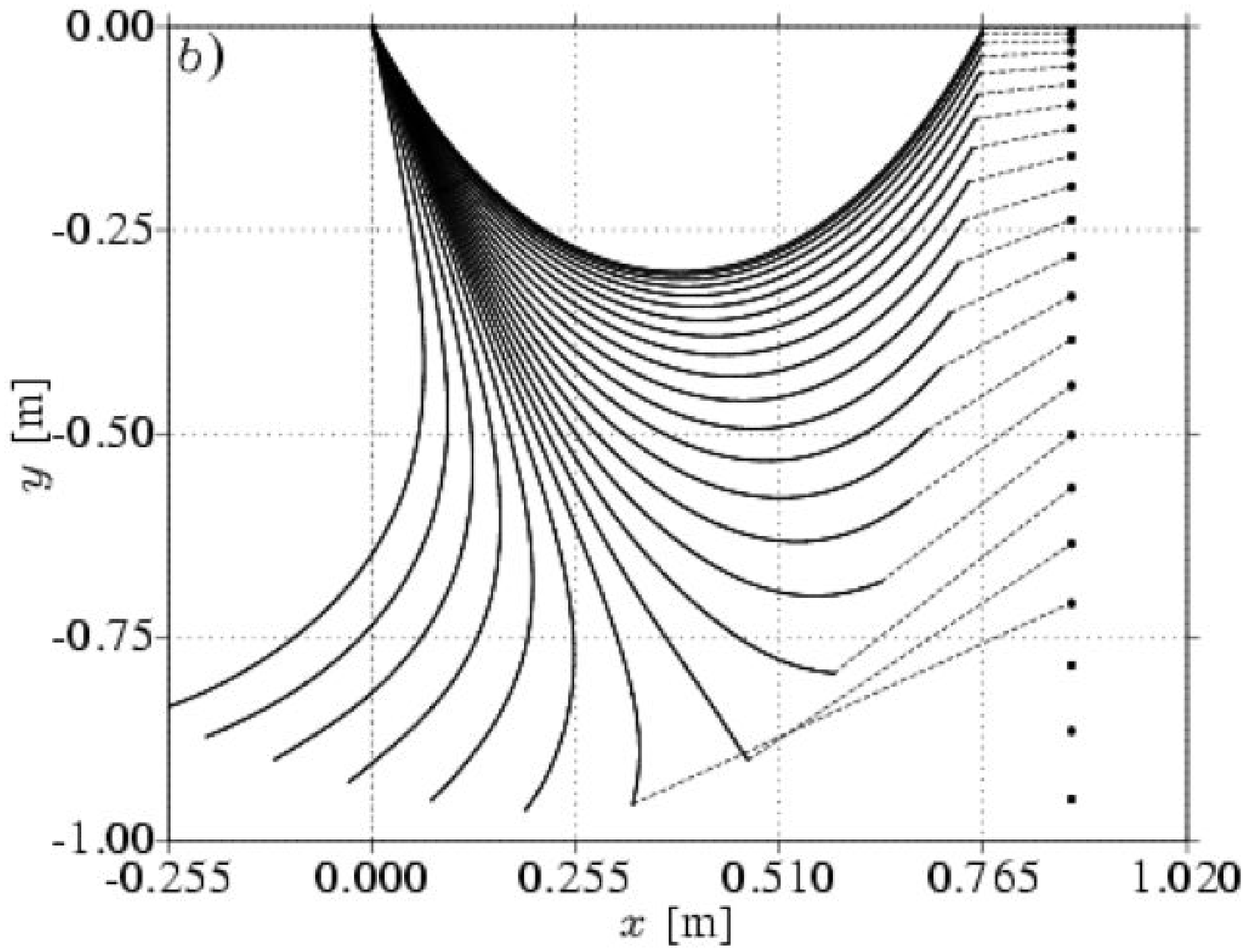}
		\includegraphics[scale=0.30,angle=-90]{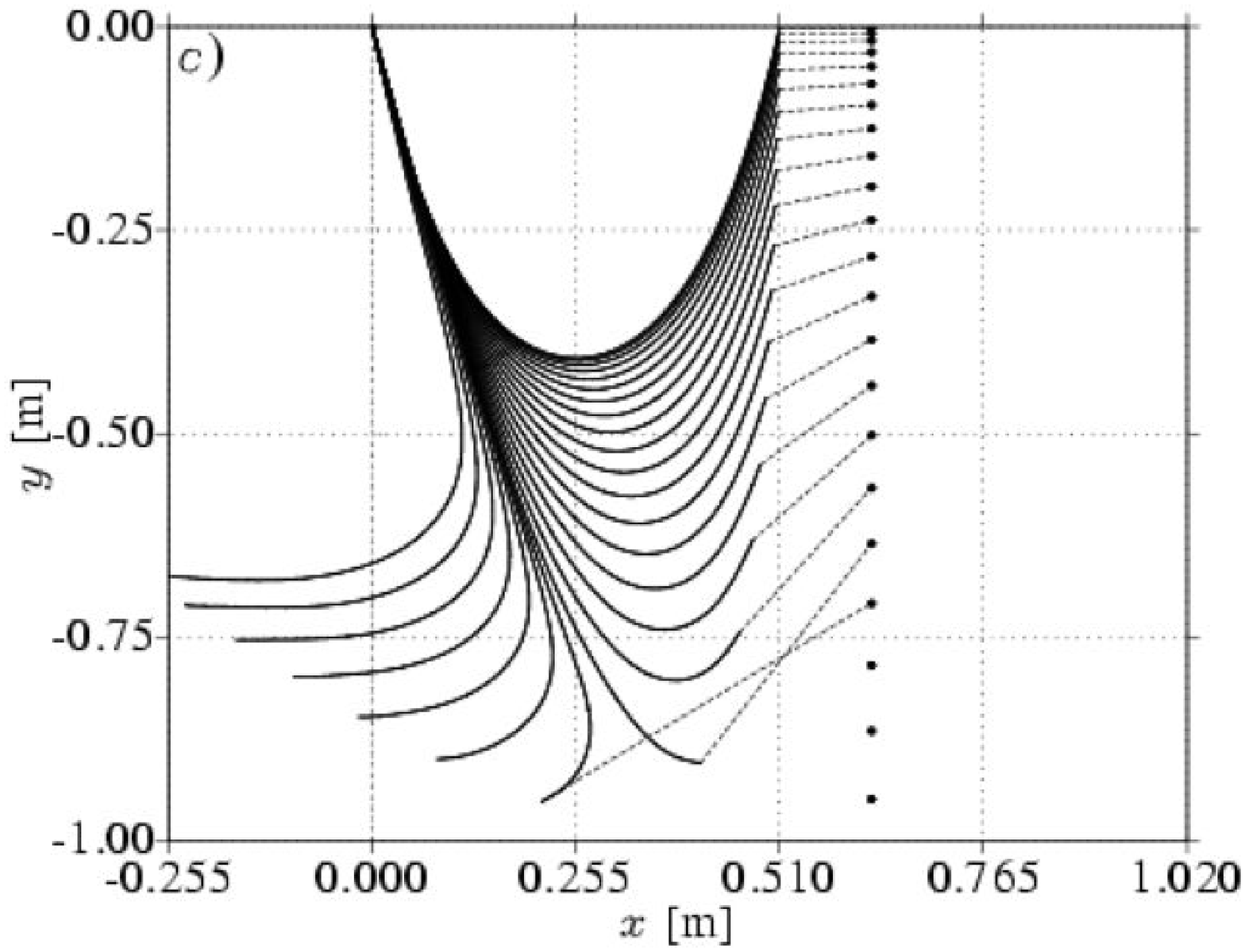}
		\includegraphics[scale=0.30,angle=-90]{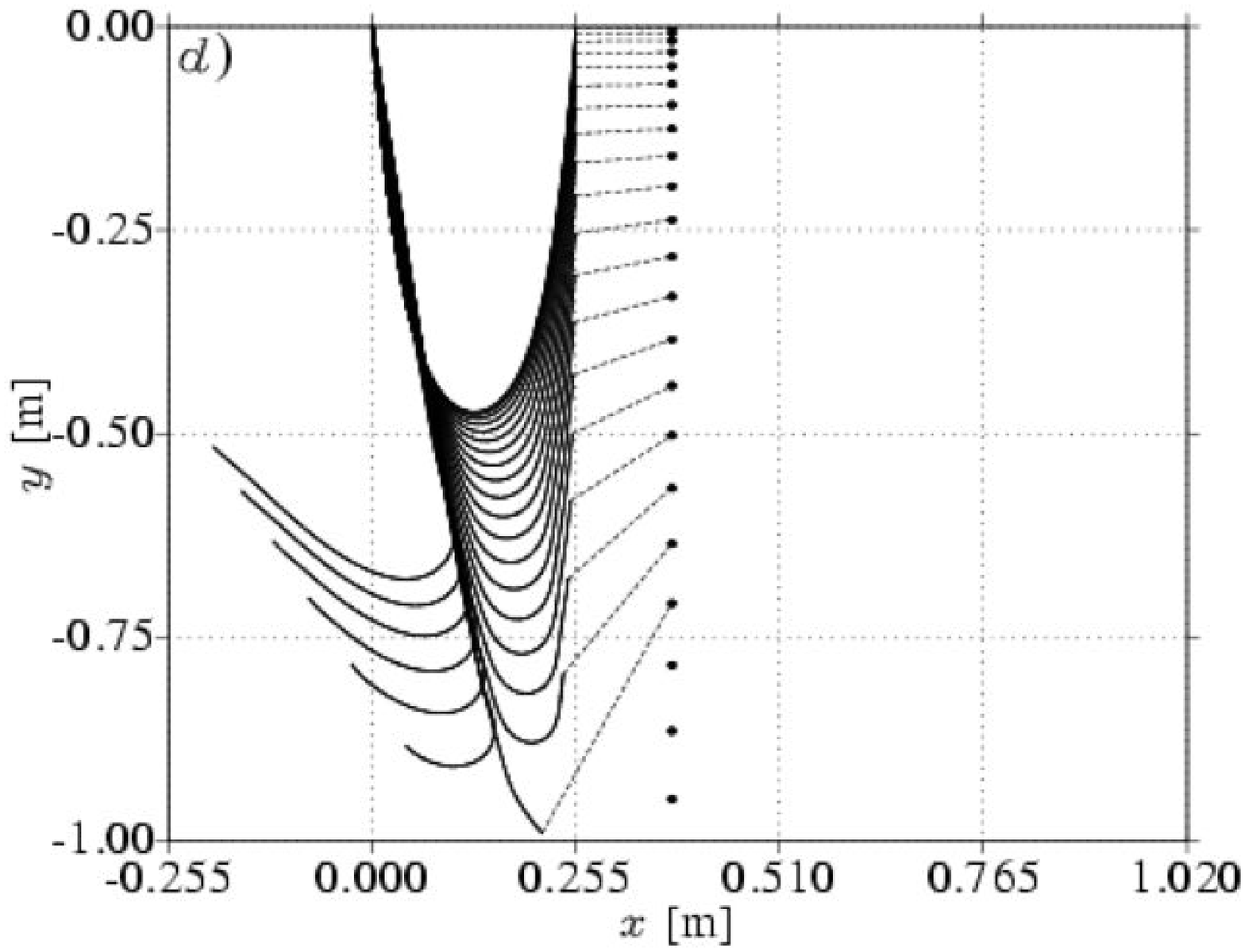}
	\end{center}
	\caption{\label{Figure009}Successive conformations of the falling chain vs. time found in numerical simulations. Simulations were performed with $n=229$, $L=1.02$~m, $M=0.0208$~kg, $g=9.81$~m/s$^2$ and the values of $r$ given in table \ref{tab1}. The initial conformations of the chain were discrete catenary curves with $a$) $x_0=1.0195$~m, $b$) $x_0=0.765$~m, $c$) $x_0=0.51$~m and $d$) $x_0=0.255$~m. Positions of the freely falling compact body are shown at the right parts of the figures; dotted lines connect them with the respective positions of the tip of the falling chain.}
\end{figure}

Quantitative analysis of the digital images recorded in the laboratory experiments provided us with sets of discrete data representing the vertical, $h$, and horizontal, $w$, fall distances of the chain tip versus time. As described above, using the data we have found the values of the dissipation parameter at which numerical simulations fit best the experimental data (table \ref{tab1}). As seen in figure \ref{Figure008} the agreement is very good. Thus, to analyze the details of the falling chain dynamics we shall be using the data obtained at small time steps from the numerical simulations.

First of all, let us analyze the most interesting question of the relation between the time dependences of the vertical fall distances of the chain tip and the compact body (figure \ref{Figure008}).

It is worth noticing that in the case $a$), where the initial conformation of the chain is straight and horizontal, the vertical fall of the chain tip and the fall of the compact body are identical up to the moment of time at which having reached its maximum vertical fall distance the tip starts moving upwards. That it should be like that becomes clear when one notices that during the fall the end part of the chain remains horizontal - its vertical motion must be thus identical with the fall of the compact body. Why the end part of the chain remains horizontal is also clear. This happens because the chain displays no elasticity and no energy is stored in its bent regions. This phenomenon, noticed in the laboratory experiments and confirmed in the numerical simulations, suggests the existence of an approximate analytical treatment. So far we have not been able to find it.

In cases $b$), $c$) and $d$) the vertical fall distance of the chain tip, up to the moment of time $t_{h_{max}}$ at which vertical fall distance of the chain tip reaches its maximum value $h_{max}$, is seen to be always ahead the vertical fall distance of the compact body. This observation is sometimes summarized by the general statement, that \textit{the chain falls faster than a compact body}.

The next question that we shall analyze is the time dependencies of the velocity $v_c$ and the acceleration $a_c$ of the chain tip. In order to do so we perform a series of numerical experiments with $x_0$ from a range $[0.1,\ldots,1.02)$~m. All parameters of the numerical simulation are the same as defined in previous section. The plot starts at $x_0=0.1$~m since because of the finite length of the chain segments at smaller initial separations the simulated dynamics of the chain fall becomes very complex. Similar effects are observed in the laboratory experiments.

\begin{figure}[htbp]
	\begin{center}
		\includegraphics[scale=0.35,angle=-90]{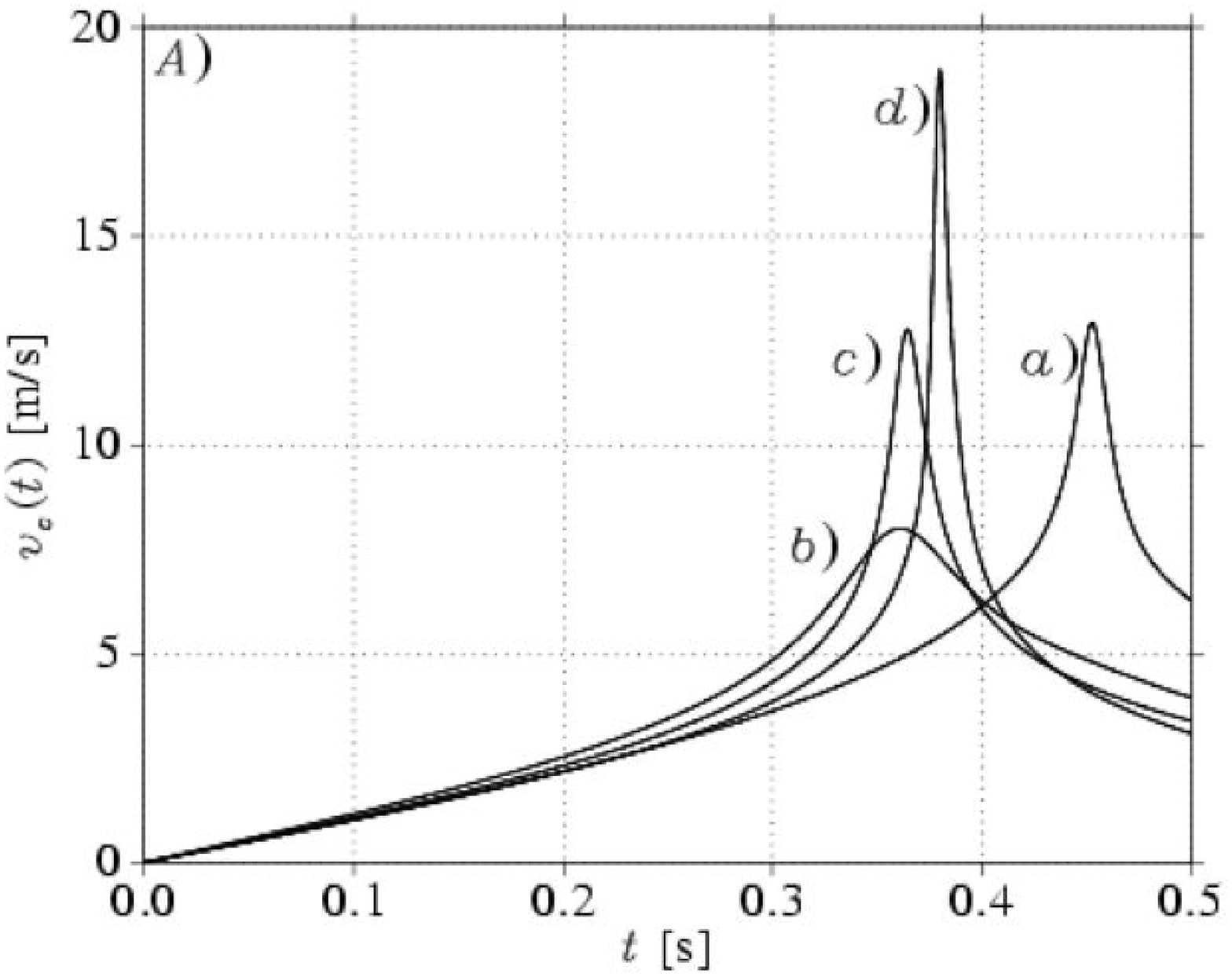}
		\includegraphics[scale=0.35,angle=-90]{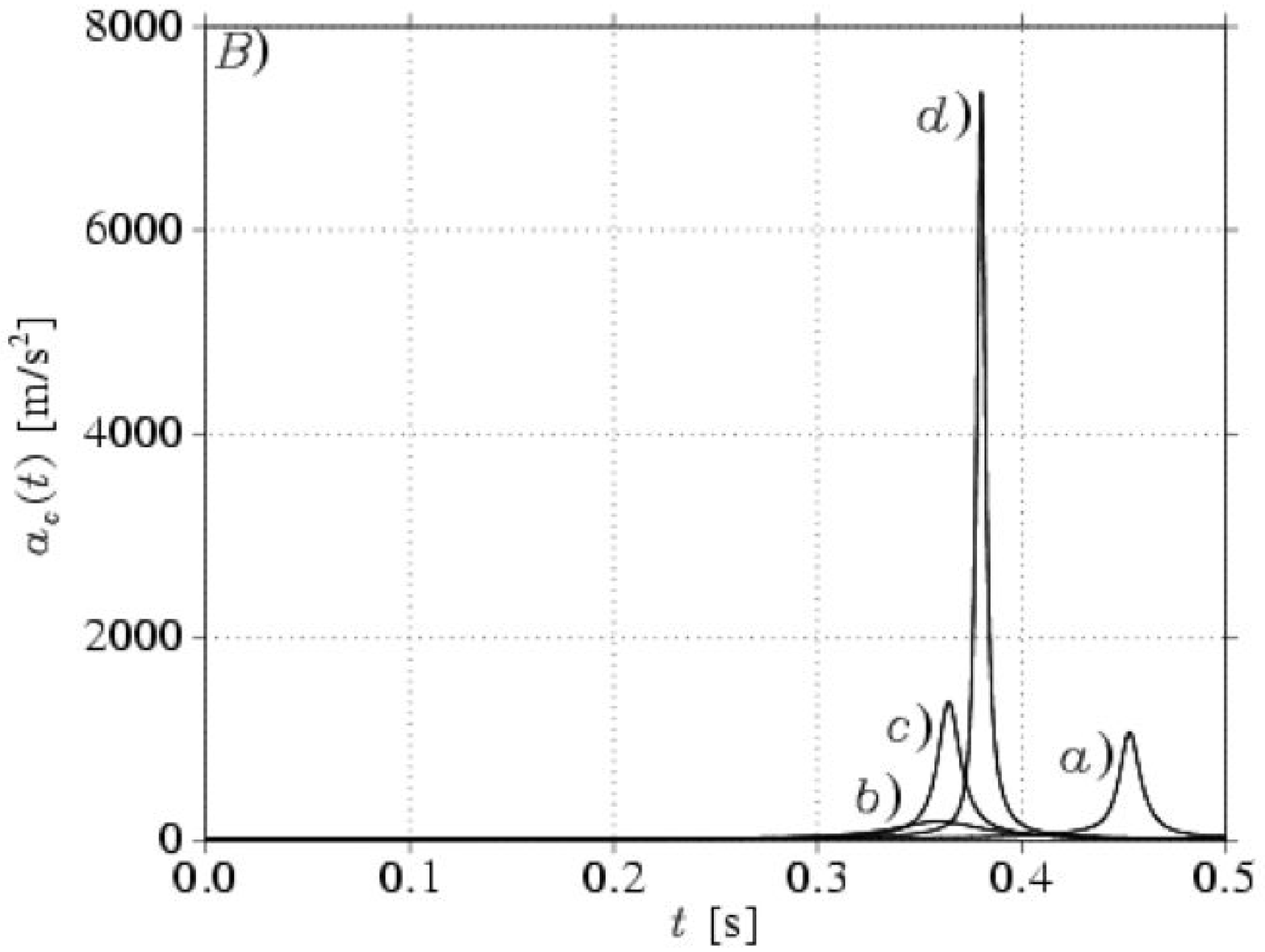}
	\end{center}
	\caption{\label{Figure010}The moduli of the velocity and the acceleration of the falling chain tip - numerical approximation of the experimental data. Initial separation of the chain ends: $a$) $x_0=1.0195$ m, $b$) $x_0=0.765$ m, $c$) $x_0=0.51$ m, $d$) $x_0=0.255$ m.}
\end{figure}

\begin{figure}[htbp]
	\begin{center}
		\includegraphics[scale=0.35,angle=-90]{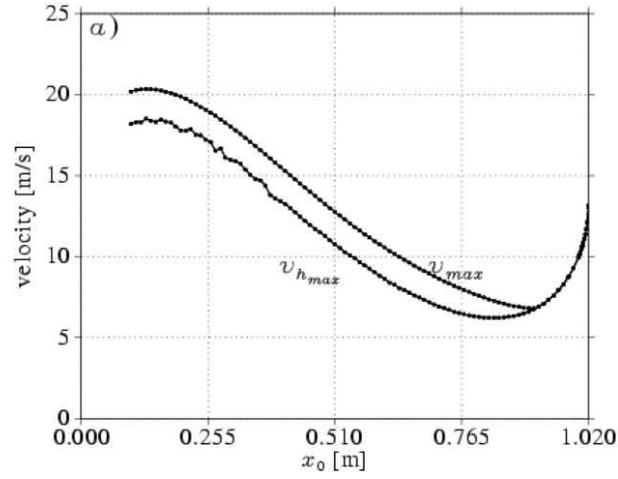}
		\includegraphics[scale=0.35,angle=-90]{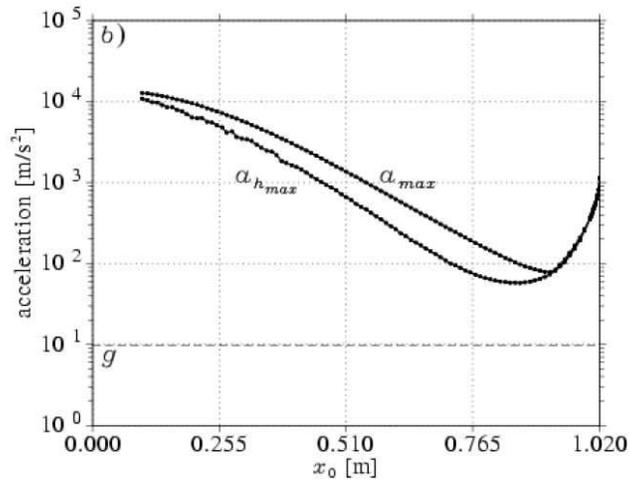}
	\end{center}	\caption{\label{Figure011} Moduli of the velocity $a$) and acceleration $b$) of the chain tip versus the initial horizontal separation of the chain ends. $v_{max}$ and $a_{max}$ are, respectively, the maximum velocity and acceleration reached by the chain tip during its fall. $v_{h_{max}}$ and $a_{h_{max}}$ are the velocity and acceleration of the chain tip observed at the moment of time at which the tip reaches its lowest position. Picture $b$ is plotted in logarithmic scale. Gravitational acceleration $g$ is marked with a dashed line.}
\end{figure}

\begin{figure}[htbp]
	\begin{center}
		\includegraphics[scale=0.35,angle=-90]{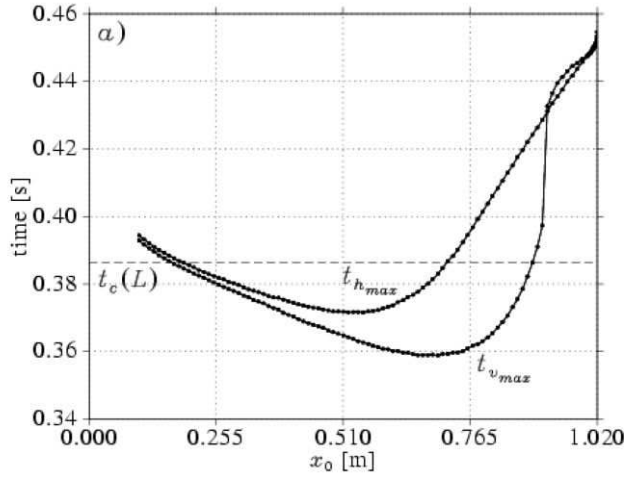}
		\includegraphics[scale=0.35,angle=-90]{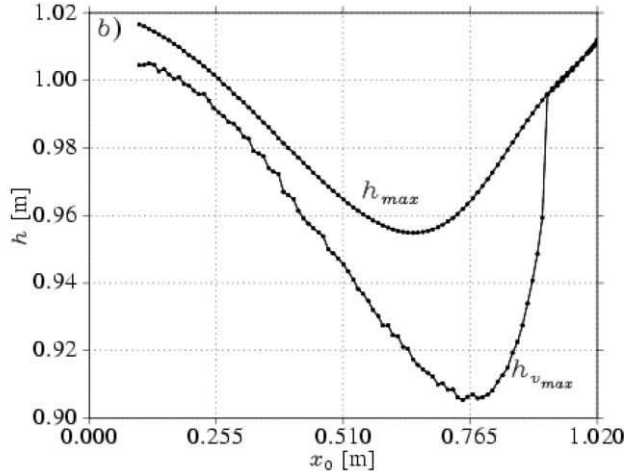}
	\end{center}
	\caption{\label{Figure012}$a$) $t_{h_{max}}$ - the time at which the chain tip reaches its lowest position; $t_{v_{max}}$ - the time at which it reaches its maximum velocity $v_{max}$. The dashed line represents the time $t_c(L) \approx 0.386722$ at which the velocity diverges in the analytical model considered in section 2. $b$) $h_{max}$ - the largest vertical fall distance reached by the chain tip; $h_{v_{max}}$ - the vertical fall distance of the chain tip at which it reaches its maximum velocity.}
\end{figure}

By velocity and acceleration we mean here the moduli of the velocity and acceleration vectors. Figure \ref{Figure010} presents both the variables versus time. As one can see, plots of the velocity versus time display distinct peaks. It seems interesting to check how high the peaks are (i.e. which the maximum velocities of the chain tips are for different initial conformations of the chain) and at which times they are reached. It seems also interesting to ask for which initial separation of the chain ends the velocity peak is highest. Answers to these questions can be found out analyzing figures \ref{Figure011} and \ref{Figure012}.

As well visible in figure \ref{Figure011}$a$, in accordance with expectations, the peak velocity value $v_{max}$ becomes largest when the initial separation of the chain ends is smallest (i.e. when the chain is maximally folded). On the other hand, contrary to expectations, the velocity peak is not smallest at the maximum $x_0=L$ initial separation but earlier, at $x_0 \approx 0.9040$~m.

Figure \ref{Figure012}$a$ reveals an interesting fact. The moment of time $t_{v_{max}}$ at which the velocity of the chain tip reaches its maximum value precedes in general the moment of time $t_{h_{max}}$ at which the chain tip reaches its maximum vertical fall distance $h_{max}$. A reverse rule is observed only at the largest initial separations $x_0$ of the chain ends.

It seems interesting to check how the time $t_{h_{max}}$ at which the chain tip reaches its lowest position depends on the initial separation $x_0$ of the chain ends. The dependence is plotted in figure \ref{Figure012}$a$ revealing that the lowest position of the chain tip is reached fastest when $x_0\approx0.5500$~m (i.e. when the initial horizontal distance between the chain ends is approximately half of its total length). $t_{h_{max}}$ proves to be longest in the case when the chain is initially straight.

\begin{figure}[htbp]
	\begin{center}
		\includegraphics[scale=0.35,angle=-90]{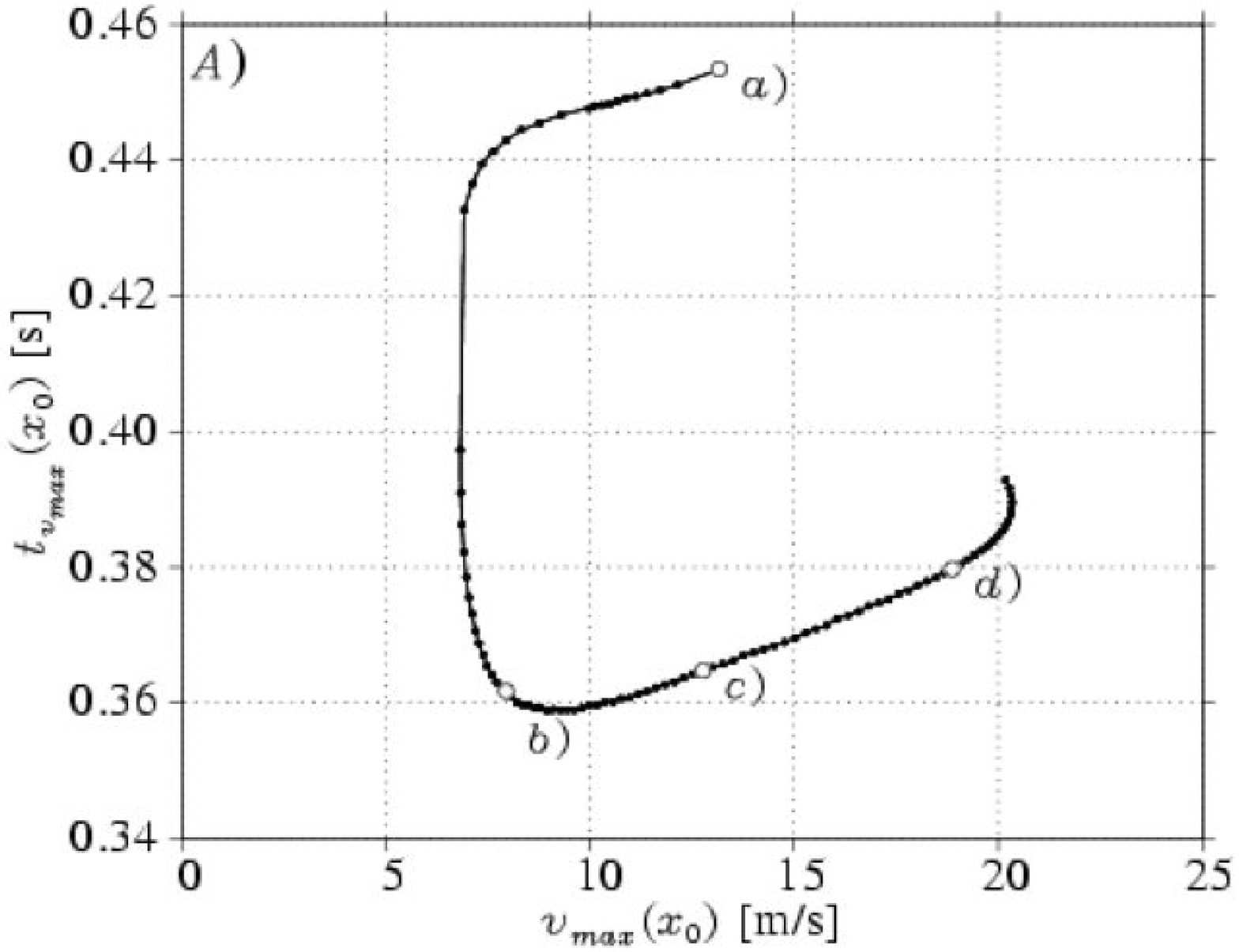}
		\includegraphics[scale=0.35,angle=-90]{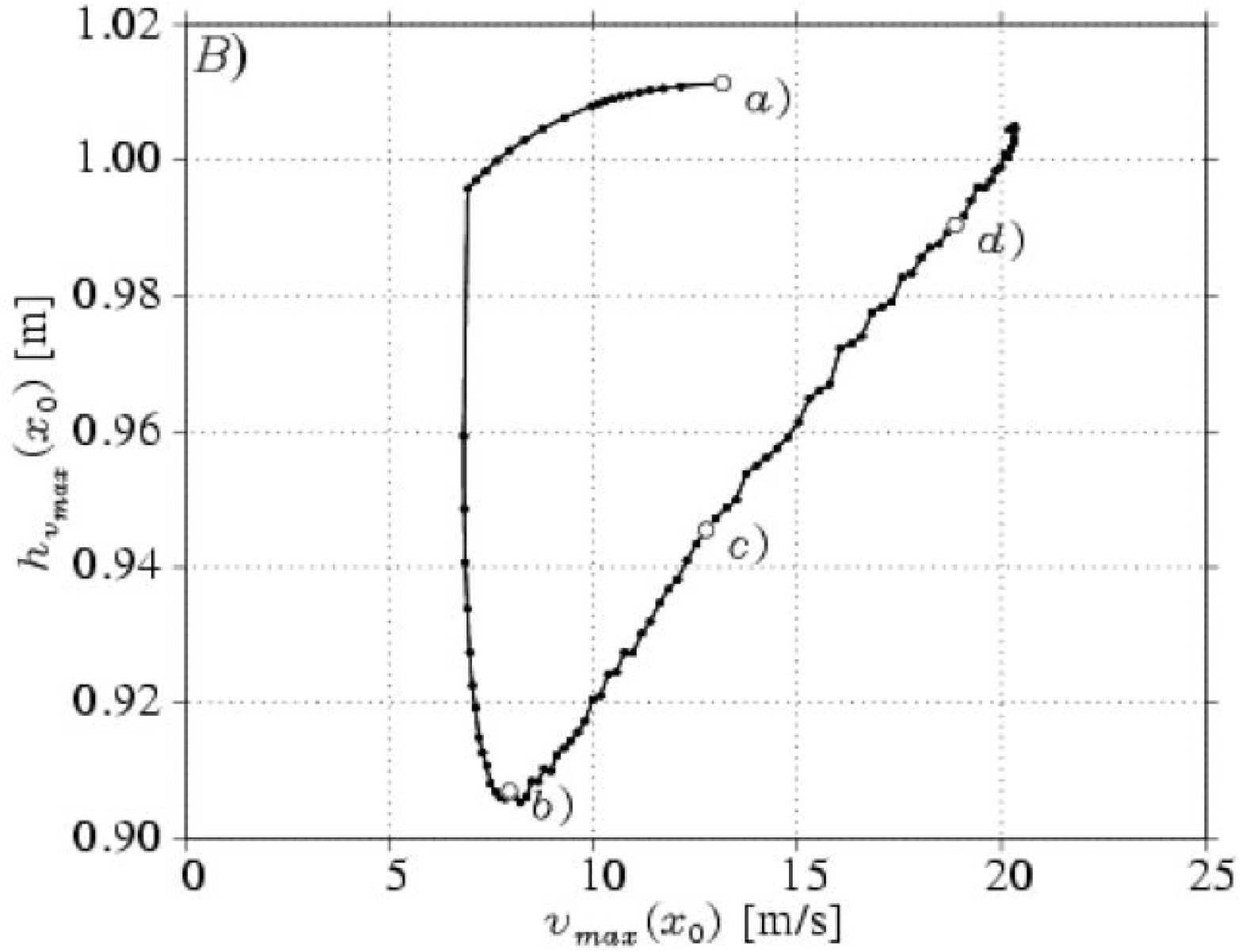}
	\end{center}
	\caption{\label{Figure013}$a$) Parametric plot of the time $t_{v_{max}}$ at which the velocity of the chain reaches its maximum velocity $v_{max}$ versus the value of the latter. $b$) Parametric plot of the maximum vertical fall distance $h_{max}$ versus the maximum velocity $v_{max}$ reached by the end tip. Bigger circles indicate the points at which the values of the plotting parameter $x_0$ (i.e. the initial separation of the chain ends) are identical to those applied in laboratory experiments.}
\end{figure}

The last question we asked analyzing the velocity data was the correlation between the value of the peak velocity $v_{max}$ and the time at which it is reached. This correlation is presented it figure \ref{Figure013}$a$. As we have demonstrated before, the peak of the velocity is highest at the smallest initial separation of the chain ends, but one should not draw the conclusion that it is thus reached in the shortest time. As seen in the figure the initial separation of the chain ends at which the velocity peak is reached fastest amounts to about $x_0\approx0.7000$~m. See also figure \ref{Figure012}$a$.

Now, let us analyze the behavior of the acceleration (figure \ref{Figure010}$b$). As in the case of the velocity plots, we also observe here clear peaks. This time, however, they are distinctly different in their height. Figure \ref{Figure011}$b$ demonstrates clearly, that the highest peak in acceleration is also observed at the smallest initial separation of the chain end. Its value, at the experimentally studied case of $x_0=0.255$~m amounts to $7352$~m/s$^2$, thus it is about 40 times larger than the value observed at $x_0=0.765$~m ($186.3$~m/s$^2$). That such large values of the acceleration are realistic was demonstrated by Krehl et. al. \cite{KrehlEngemannSchwenkel1998} who studied the dynamics of the cracking whip. Figure \ref{Figure011}$b$ demonstrates that accelerations at the lowest positions of the chain tip are not the maximum ones.

It seems interesting to check the relation between the time $t_{v_{max}}$ at which the chain tip reaches its maximum velocity and the time $t_{h_{max}}$ at which it reaches its maximum fall distance. Figure \ref{Figure012}$a$ presents the relation. As one can see, except for the largest values of the initial separation $x_0$ the maximum velocity is reached before the chain tip reaches its maximum fall distance.

Figure \ref{Figure012}$b$ presents an answer to a differently posed question. We ask about the relation between the maximum fall distance of the chain tip $h_{max}$ and the fall distance $h_{v_{max}}$ at which the tip reaches its maximum velocity. As before one can clearly see that in general the maximum velocity is reached before the chain reaches its maximum fall distance. 

Parametric relations between a) $t_{v_{max}}$ and $v_{max}$, b) $h_{max}$ and $v_{max}$ found at given values of the initial separation $x_0$ are plotted in figures \ref{Figure013}$a$ and \ref{Figure013}$b$. As well visible in the figures the range of large $x_0$ proves to be very interesting. Small changes of $x_0$ lead here to large changes of $t_{v_{max}}$ and $h_{max}$. This range of $x_0$ needs a special attention.

\section{Summary and discussion}

A chain falling in the gravitational field can be seen as a model of other systems such as the cracking whip \cite{GorielyMcMillen2002,GorielyMcMillen2003}. At the first sight the above statement may seem not true, since in the cracking whip problem, the gravitational forces are in general neglected. Let us notice, however, that the end of the folded whip attached to the whip handle is subject to a strong acceleration. Changing the laboratory reference frame to the non-inertial frame moving with the end of the handle, we introduce into the system strong inertial forces equivalent to the gravitational ones. This explains the validity of the initial remark.

Experiments we performed revealed a few new, interesting facts concerning the dynamics of the falling chain. Let us summarize them.

\begin{enumerate}
	\item {Both the velocity and acceleration dependencies on time display distinct peaks the height of which depends on the initial separation of the chain ends. The highest peaks are observed for smallest initial separation. There exists an approximate, analytical description of tightly folded chain dynamics explaining the origin of the rapid increase of the velocity and acceleration. The theory is however unable to predict the finite height of the peaks. (In the analytical model both the velocity and the acceleration diverge.) 
}
	\item {The velocity peak is observed to be reached fastest for initial separation $x_0=0.6863~L$, where $L$ is the length of the chain, whereas its amplitude is smallest for $x_0=0.8863~L$.
}
	\item {It seems very interesting that in the case in which the initial separation of the chain ends is largest, the dynamics of the vertical fall of the chain tip proves to be identical with the dynamics of the fall of a compact body. That is should be the case becomes obvious when one notices that the end part of the chain remains horizontal during the fall. This observation suggests the existence of an approximate analytical treatment. It is not known yet.
}
	\item {As a rule, the time at which the chain tip reaches its maximum velocity generally comes before the time at which it reaches its lowest vertical position. Only at the initial separation of the chain ends larger than $0.8863~L$ to $0.9608~L$ the rule is reversed.
}
	\item {The ratio between the the largest and smallest acceleration peaks is about $166.5$, which is unexpectedly large. This may have some practical implications since at the times, when the acceleration reaches its highest values, forces acting on the chain tip also become very large what may lead to a damage of the chain.
}
\end{enumerate}

Dynamics of the falling chain hides certainly a few more interesting details. The same, even to a larger extent, concerns the dynamics of the falling rope. In the latter case the dissipation plays a much more important role and elasticity becomes a crucial factor. Laboratory and numerical experiments are waiting to be carried out.

\newpage
\section*{Tables}

\begin{table}[htbp]
\caption{Distance between experimental and numerical results}
\begin{center}
\begin{tabular}{ccccccc} \hline \hline
\label{tab1}
Experiment: & \vline & $\delta$~[m] \\ \hline
$a$) & \vline & $0.007672$\\
$b$) & \vline & $0.006964$\\
$c$) & \vline & $0.005912$\\
$d$) & \vline & $0.004552$\\
\hline \hline
\end{tabular}
\end{center}
\end{table}

\newpage
\section*{Figure Captions}

\newpage

\newpage

\newpage

\newpage

\newpage

\newpage

\newpage

\newpage

\newpage

\newpage

\newpage

\newpage

\end{document}